\documentstyle[12pt]{article}

\def\a{\alpha }

\def\L{\Lambda }

\begin{document}
\begin{flushright}
CERN-TH.7160/94
\end{flushright}
\begin{flushright}
JINR E2-94-45
\end{flushright}

\vskip 0.15in
\begin{center}

{\bf \LARGE
 The Jacobi Polynomials QCD analysis of the
 CCFR Data for $xF_3$ and
the $Q^2$-Dependence of the
Gross--Llewellyn Smith  Sum Rule
%: the Problems of the  Comparison with the Theoretical QCD Predictions.
          }

\end{center}
\vskip 0.1in
\begin{center}

{\bf Andrei L. Kataev}
\\Theory Division, CERN, CH-1211 Geneva 23, Switzerland;\footnote{E-mail:
KATAEV@CERNVM.BITNET}
\\
 Institute for Nuclear Research of the Russian Academy of Sciences,
Moscow 117312, Russia\footnote{On leave of absence at CERN
from February 1994}
\\{\bf Aleksander V. Sidorov}
\\
Bogoliubov Theoretical Laboratory, Joint Institute for Nuclear Research,
141980 Dubna, Russia\footnote{E-mail: SIDOROV@THEOR.JINRC.DUBNA.SU}

\end{center}
\vskip 0.3in
\begin{center}
{\bf \Large Abstract}
\end{center}
\vskip 0.1in

We present the results of our QCD analysis of the recent
CCFR data for the structure function $xF_3 (x,Q^2)$  of the
deep-inelastic neutrino--nucleon scattering.
The analysis is based on the Jacobi polynomials expansion of
the structure functions. The concrete results for the parameter
$\Lambda_{\overline {MS}}^{(4)}$ and the shape of quark distributions
are determined.
At the reference scale $|Q_0^2|$=3 $GeV^2$ our results are in
satisfactory  agreement with
the ones obtained by the CCFR group with the help of another method.
The $Q_0^{2}$-dependence of
 the experimental data for the Gross--Llewellyn Smith  sum rule
is extracted in  the wide region of high-momentum transfer.
Within systematical experimental uncertainties the results obtained
are consistent with the perturbative QCD predictions.
We reveal the effect of the discrepancy  between
our results  and the analysed perturbative QCD predictions
at the level of the statistical error bars.
The importance of taking  account, in our procedure, of a
still unknown next-next-to-leading approximation of the
moments of the structure function $xF_3 (x,Q^2)$ is stressed.

\vskip 1cm
\noindent   CERN-TH.7160/94\\
\noindent   JINR E2-94-45\\
\noindent  February 1994
\addtocounter{page}{-1}
\thispagestyle{empty}
\vfill\eject
\pagestyle{empty}
% empty page for printing shop
\clearpage\mbox{ }\clearpage
\pagestyle{plain}
\setcounter{page}{1}

\newpage
\setcounter{equation} {0}

{\bf 1. Introduction}

The deep-inelastic lepton--nucleon scattering is the
source of  important information about the nucleons structure.
In the last years the accuracy of the obtained experimental data became
large enough to study in detail the status of the comparison of the available
data with the theoretical predictions of QCD in the different regions of
 momentum transfer \cite{Altarelli}.

The most precise data for the structure function (SF) $xF_3 (x,Q^2)$  was
recently obtained by the CCFR collaboration at the FERMILAB collider
\cite{prep1}--\cite{prep3}
(for a detailed description see Refs. \cite{PHD}).
The theoretical analysis of the obtained experimental data for the process
of  nucleon destruction by the charged currents was made by the members
of the CCFR collaboration, with the help of the computer
program developed
in Ref. \cite{prog} and based on the direct integration of the
Altarelli--Parisi equation \cite{AP}.
The fits to the data \cite{prep1}--\cite{prep3} were only
made where perturbative
QCD is expected to be valid. The results of a next-to-leading order
(NLO)
fit of the non-singlet SF $xF_3 (x,Q^2)$ were obtained
for the different values of the $Q^2$ cut. In particular in the case of the
cut $|Q^2|>10\ GeV^2$ the CCFR collaboration got the following value of
the
QCD scale parameter $\Lambda_{\overline{MS}}^{(4)}$ (see \cite{prep3}) :

\begin{equation}
\Lambda_{\overline{MS}}^{(4)}=171 \pm 32 (stat.) \pm 54 (syst)\ MeV\ ,
\label{La}
\end{equation}
which corresponds to $f=4$ numbers of active flavours. Notice that the
former cut allows one to neglect the effects of the high-twist (HT)
contribution
and the target mass (TM) corrections to $xF_3 (x,Q^2)$.

Another important characteristic of the deep-inelastic neutrino--nucleon
scattering is the Gross--Llewellyn Smith (GLS) sum rule \cite{GLS}
\begin{equation}
GLS (Q^2)=\frac{1}{2}\int_{0}^{1}\frac{xF_{3}^{\overline{\nu}p+{\nu}p}(x,Q^2)}
{x}dx .
\label{gls}
\end{equation}
In the work of Ref.
\cite{prep2}, the following result of the measurement of the GLS sum
at the scale $|Q_0^2|=3\ GeV^2$ was reported :
\begin{equation}
GLS(|Q_0^2|=3\ GeV^2)=2.50\pm 0.018 (stat)\pm 0.078 (syst) .
\label{result}
\end{equation}
This result has been obtained by a special procedure of either interpolation or
extrapolation of the $xF_3 (x,Q^2)$ data with the help of the best fit of
the $Q^2$-dependence and taking into account the logarithmic $Q^2$-dependence
as predicted by QCD \cite{prep2,PHD}. Equation \ref{result}
was already used
as the bases of the NLO QCD analysis \cite{ChK,Ross} by
taking into account
the NLO perturbative QCD corrections to the GLS sum rule \cite{GL}
(which were confirmed in \cite{vanNeerven}) and the
corresponding HT
corrections \cite{SV} estimated by the QCD sum rules method in
Refs. \cite{BK,Ross}. However, it should be stressed that the explicit
$Q_0^2$-dependence of the CCFR data for the GLS sum rule remained
non-investigated.

In this work this important problem is studied with the help of the method
of the SF reconstruction over their Mellin moments, which
is based on the expansion of the SF over the
Jacobi polynomials  \cite{Jacobi}. This method was
developed \cite{Barker,Kriv} (see also \cite{ChR}),
 discussed \cite{LS} and previously used
by the BCDMS collaboration in  concrete physical applications
\cite{BCDMS} (see also \cite{Kotikov}).\\[2mm]

{\bf 2. The Method of the QCD Analysis of SF}

We  first recall the basic steps of the method used in this work
to  reconstruct of SF in  the $x$-representation
over their Mellin moments.

Let us define the Mellin moments for the non-singlet SF $xF_3 (x,Q^2)$ :
\begin{equation}
M_n^{NS}(Q^2)=\int_{0}^{1}x^{n-1}F_{3}(x,Q^2)dx ,
\label{mom}
\end{equation}
where
$n$ =2, 3, 4, ... .\footnote{Note that the first moment $M_1^{NS}(Q^2)$
is nothing more than the GLS sum rule defined in Eq. (\ref{gls}).}
 The $Q^2$-evolution of the moments  is  given
by the solution of the corresponding renormalization-group equation.
In the NLO approximation of perturbative QCD it can be presented in the
following form \cite{Yndurain}:
\begin{equation}
M_{n}^{NS}(Q^2)
 =\left [ \frac{\alpha _{s}\left ( Q_{0}^{2}\right )}
{\alpha _{s}\left ( Q^{2}\right )}\right ]^{d_{n}}
H_{n}\left (  Q_{0}^{2},Q^{2}\right ) M_{n}^{NS}(Q_{0}^2),
\label{e1}
\end{equation}
where  $d_{n}=-\gamma_{NS}^{0}/2\beta_0$ and
\begin{equation}
H_{n}\left (  Q_{0}^{2},Q^{2}\right ) =
\frac{1+C_{NS}^{1}(n)\frac{\alpha_{s}\left( Q^{2}\right )}{4\pi }}
{1+C_{NS}^{1}(n)\frac{\alpha _{s}\left ( Q_{0}^{2}\right )}{4\pi }}
 {\left [ \frac
      {1+\beta _{1}\frac{\alpha _{s}\left ( Q^{2}    \right )}
                                              {4\pi \beta_{0}} }
      {1+\beta _{1}\frac{\alpha _{s}\left ( Q_{0}^{2}\right )}
                                              {4\pi \beta_{0}} }
              \right ] }^{p(n)}
\label{e2}
\end{equation}
\begin{equation}
p(n) = \frac{1}{2}\left [ \frac{\gamma_{NS}^{(1)}}{\beta_{1}} -
\frac{\gamma_{NS}^{(0)}}{\beta_{0}}\right ] .
\nonumber
\end{equation}
The NLO approximation of the QCD coupling constant $\a_s(Q^2)$ can be expressed
through the scale parameter $\L_{\overline{MS}}$ as
\begin{equation}
\frac{\a_s(Q^2)}{4\pi}= \frac{1}{\beta_0 \ln(Q^2/\L_{\overline{MS}}^2)}-
\frac{\beta_1 \ln\ln(Q^2/\L_{\overline{MS}}^2)}{\beta_0^3
\ln^2(Q^2/\L_{\overline{MS}}^2)}
\label{alnlo}
\end{equation}
where  $\beta_0$ and $\beta_1$, namely
\begin{equation}
\beta_0=11-\frac{2}{3}f~~~and~~~ \hspace{10mm} \beta_1=102-\frac{38}{3}f,
\nonumber
\end{equation}
are the leading-order (LO) and NLO coefficients
of the QCD $\beta$-function, which was originally calculated at the
next-next-to-leading order (NNLO) level \cite{TVZ} and confirmed in
Ref. \cite{LarVer}.

The analytic expressions for the LO coefficient $\gamma_{NS}^{(0)}$
of the anomalous dimension function of the non-singlet operator
and the corresponding expression
for the NLO coefficient
$\gamma_{NS}^{(1)}$ can be found, e.g. in the textbook
of Ref. \cite{Yndurain}.
 For the neutrino--nucleon deep-inelastic
scattering, the NLO coefficient $C_{NS}^{1}(n)$ of
the coefficient function
is known from the results of Ref. \cite{Bardeen}.

 Following the methods of \cite{Jacobi}-\cite{ChR}, one can
expand the SF in the set over Jacobi polynomials $\Theta_n ^{\a , \beta}(x)$  :
\begin{equation}
xF_{3}^{N_{max}}(x,Q^2)=x^{\a}(1-x)^{\beta}\sum_{n=0}^{N_{max}}
a_{n}(Q^2) \Theta_n ^{\a , \beta}(x), \label{e5}
\end{equation}
where $N_{max}$ is the number of polynomials and $a_{n}(Q^2)$
are the  coefficients of the corresponding expansion.

The Jacobi polynomials  $\Theta_n ^{\a , \beta}(x)$ obey the orthogonality
relation
\begin{equation}
\int_{0}^{1}dxx^{\a}(1-x)^{\beta}\Theta_{k} ^{\a , \beta}(x)
\Theta_{l} ^{\a , \beta}(x)=\delta_{k,l}\ ,
\label{e8}
\end{equation}
and can be expressed as the series in  powers of $x$:

\begin{equation}
\Theta_{n} ^{\a , \beta}(x)=
\sum_{j=0}^{n}c_{j}^{(n)}{(\a ,\beta )}x^j ,
\label{e9}
\end{equation}
where $c_{j}^{(n)}{(\a ,\beta )}$ are the coefficients that
 expressed through $\Gamma$-functions.

Using now Eqs. (\ref{e8}), (\ref{e9}) and (\ref{mom}), one can relate
the SF with its  Mellin moments
\begin{equation}
xF_{3}^{N_{max}}(x,Q^2)=x^{\a}(1-x)^{\beta}\sum_{n=0}^{N_{max}}\Theta_n ^{\a ,
\
(x)\sum_{j=0}^{n}c_{j}^{(n)}{(\a ,\beta )}
M_{n}^{NS} \left ( Q^{2}\right ),   \\
\label{e7}
 n = 2,3, ...  \nonumber
\end{equation}

The relations of
Eqs. (\ref{e1}), (\ref{e2}) and (\ref{e7})  form the
basis of the computer program created by the authors of Ref. \cite{Kriv}.
It  was previously tested and used
by the members of the BCDMS collaboration in the
course of detailed QCD analysis of the experimental data for the SF of the
deep-inelastic muon--nucleon scattering \cite{BCDMS,LS}.
\\[2mm]

{\bf 3. The  Procedure of the QCD Fit of  $xF_3$ Data}

In accordance with the original non-singlet fit of the CCFR collaboration
\cite{prep2,PHD} in the proces of the studies of their experimental data,
we choose the parametrization of the parton distributions at  fixed
momentum transfer $Q_0^2$ in the simplest form :
\begin{equation}
  xF_{3}(x,Q_0^2)=Ax^{b}(1-x)^{c} ,
\label{e10}
\end{equation}
which was originally used by the CCFR collaboration to get the
result of Eq. (\ref{result}) for the GLS sum rule.

The constants $A$, $b$, $c$ in
Eq. (\ref{e10}) and the QCD scale parameter
$\Lambda$ are considered as  free parameters, which should be determined
for  concrete values of $Q_0^2$.
The values of the parameters $A$, $b$ and $c$
depend on the value of $Q_0^2$.
In order to avoid the influence of the HT effects and the TM
corrections, we  use the experimental points of the concrete
CCFR data \cite{PHD} in the plane
$(x,Q^2)$ with $0.015<x<0.65$ and
$10\ GeV^2<|Q^2|<501\ GeV^2$.

It should be stressed that for  deep-inelastic processes with charged
currents, which were dealt with in the CCFR experiment, the former region
of momentum transfer implies that there are four active flavours and that
we have to use $f=4$ in formulae (\ref{e1}),(\ref{e2}) and (\ref{alnlo}).
In view of this fact we will not take threshold effects into account
in the process of the present analysis.

We are now  ready to discuss the main steps of our analysis :
\begin{itemize}
\item
The parametrization of
Eq. (\ref{e10}) allows us to calculate the concrete
expression for the Mellin moments through $\Gamma$-functions that depend
on the parameters $A$, $b$ and $c$, namely the expression
$M_{n}^{NS}(Q_0^2, A,b,c)$.
As was estimated in \cite{Kriv}, in order to get the
accuracy better then $10^{-3}$ in the procedure of the SF
reconstructions, it is
sufficient to use in Eq. (\ref{e5}) $N_{max}=10$. However to make the
analysis even  more reliable  we will use $N_{max}=12$.

\item
The next step is to use the QCD theoretical evolution of
Eqs. (\ref{e1}) and  (\ref{e2}) for
the calculation of each  Mellin moment at $Q^2$ values
that corresponds to concrete  experimental points $Q_{exp}^2$ for the
SF $xF_3(x,Q^2)$.
At this stage the essential dependence of the Mellin moments from
the QCD scale parameter $\Lambda$  appears :
%$M_{n}^{NS}(Q_{exp}^2, A,b,c,\L)$.
\begin{equation}
M_n^{NS}(Q_0^2,A,b,c)\stackrel{QCD}{\rightarrow}
M_n^{NS}(Q_{exp}^2,A,b,c,\L),
\nonumber
\end{equation}
\item
Using now Eq. (\ref{e7}) we can  reconstruct the theoretical expression
for the SF $xF_{3}(x,Q^2)$, namely $xF_{3}^{(N_{max}=12)}(x,Q^2,A,b,c,\L)$
for all experimental points $(x_{exp},Q^2_{exp})$.
\item The numerical values of the parameters $\a$ and $\beta$ which
define the corresponding Jacobi polynomials $\Theta_n ^{\a , \beta}(x)$,
can be choosen  such as
to achieve the  fastest convergence
of the series in the r.h.s. of Eq. (\ref{e5}).
This  procedure was discussed
in Ref. \cite{Kriv}. In accordance with the results of
Ref. \cite{Kriv} we
use $\a=0.12$ and $\beta=2.0$.
\item
The determination of the free parameters of the fit (namely $A$, $b$, $c$
and $\L$) from the CCFR experimental data for $xF_{3}(x_{exp},Q_{exp}^2)$
is made by  minimization of $\chi^2$ by the MINUIT program,
which automatically calculates the statistical errors of the parameters
also.
\item
The obtained values of the parameters $A$, $b$ and $c$ depend on the
 reference scale $Q_0^2$, which enters into the
expression for $M_n^{NS}(Q_{exp}^2,A,b,c,\L)$ (and thus for \\
$xF_{3}^{(N_{max}=12)}(x,Q^2,A,b,c,\L)$)
through Eqs. (\ref{e1}) and  (\ref{e2})).
\item
In order to evaluate the numerical value of the GLS sum rule at the
reference scale $Q_0^2$, it is
necessary now to substitute  $A(Q_0^2)$, $b(Q_0^2)$
and $c(Q_0^2)$ for their concrete values
in Eq. (\ref{e10}) and to calculate the integral
of Eq. (2).
\item
Repeating the above  procedure for the different values of
$Q_0^2$, we  determine the experimental dependence of the GLS
sum rule from the momentum transfer.
\item
The described fit will be made both in the LO and NLO of perturbative
QCD. In the process of the LO fit we will use the LO approximations
of the anomalous dimension function, coefficient function and the QCD
coupling constant $\alpha_s$ defined through the corresponding scale
parameter $\L_{LO}$ as
$\a_s{(Q^2)}={4\pi}/\beta_0\ln(Q^2/\L_{LO}^2)$.
\end{itemize}
\vspace{2mm}

{\bf 4. The Results  of the QCD Fit of the CCFR $xF_3$ Data}

The fitting procedure discussed in Section 3 was applied by us to the
analysis of the CCFR data for the non-singlet SF measured in
the neutrino deep-inelastic scattering \cite{PHD}.
The results of the fit at different values of $Q_0^2$ are presented
in Table 1.

Several comments are in order:
\begin{itemize}
\item The stable value of $\Lambda$ for fits with different $Q_0^2$ both
for LO and NLO indicates the stability and the self-consistence of the
method used.
\item It is easy to see from Table 1,
that within the statistical errors,
the results of our NLO fit of the
parameter $\L_{\overline{MS}}^{(4)}$ in the wide region of $Q_0^2$
are in agreement with the result (\ref{La}), obtained by the CCFR group
with a little bit more complicated parametrization of the SF
$xF_{3}(x,Q_0^2)=Ax^{b}(1-x)^{c}+Dx^{e}$,
in the same kinematic region (see ref.\cite{PHD}). The estimation
of the systematic error in Eq. (\ref{La}) remains true for our results.
\item Our results for $\L_{\overline{MS}}^{(4)}$ are in
exact agreement with the outcome of the combined non-singlet fit of
the CCFR data for the
$xF_3(x,Q^2)$ and $F_2(x,Q^2)$ SFs \cite{prep3,PHD},
 namely
$\L_{\overline{MS}}^{(4)}=210\pm 28 (stat)\pm 41(syst)\ MeV$.
\item The inequality $\chi^{2 (LO)}_{d.f.} > \chi^{2 (NLO)}_{d.f.}$ indicates
that the NLO  is preferable,
for the description of the  experimental data.
Notice, that even if HT-effects  and the TM corrections have been
neglected,
$\chi^2_{d.f.}$ is rather good in the wide region of $Q_0^2$,
which includes even low momentum transfer.

\item The results of our fit for the parameters of
quark distributions can be compared with the results obtained by the CCFR
group with the help of another program \cite{prog} at the reference scale
$|Q_0^2|=3\ GeV^2$. This comparison is presented in Table 2.
One can see that the agreement between the NLO results is satisfactory.
\end{itemize}
\vspace{2mm}

{\bf 5. The $Q^2$-Dependence of the GLS Sum Rule VS Experiment}

We consider the results of Table 1 for the GLS sum rule as the
experimental points in the wide region of $Q_0^2$. The corresponding
statistical errors can be estimated using the statistical errors of the
parameters $A$, $b$ and $c$ of quark distributions as presented in the second
column of Table 2. Using the concrete expression for the first Mellin moment
through the quark distributions of
Eq. (\ref{e10}), we find that the statistical
errors for the GLS sum rule are within $4\%$--$5\%$. The systematical
uncertainty was determined by the CCFR group itself \cite{prep2,PHD} (see
Eq. (\ref{result})).

Taking into account these estimates of the statistical and experimental
uncertainties of the experimental outcomes  of our NLO fit, we get the
following value for the GLS sum rule  at the scale $|Q_0^2|=3\ GeV^2$ :

\begin{equation}
GLS(|Q_0^2|=3\ GeV^2)=2.446 \pm 0.100 (stat)\pm 0.078 (syst)
\label{nresult}
\end{equation}
which is in agreement with the result (\ref{result}) obtained by the
CCFR group. The smaller statistical error of the CCFR result
of Eq. (\ref{result}) comes from their more refined analysis of this
type of  experimental uncertainties.

Let us now compare the experimental behaviour of the GLS sum rule
with  the corresponding perturbative QCD  predictions for the first
Mellin moment, which determines the theoretical expression for
the GLS sum rule.
The result we are interested in has
the  form

\begin{equation}
GLS_{QCD}(Q^2)=3\left [1-\frac{\a_s(Q^2)}{\pi}
+O(\a_s^2(Q^2)) + O(\frac{1}{Q^2}) \right ].
\label{glsth}
\end{equation}
 The estimates of the HT corrections are presented in Refs.
\cite{BK,Ross} and the NLO corrections of order $O(\a_s^2(Q^2))$ and
the NNLO corrections of order $O(\a_s^3(Q^2))$ were
analytically calculated in \cite{GL,vanNeerven} and \cite{LV},
respectively.

It is worth emphasizing that putting $n=1$ in the LO QCD expression for
the moment $M_n^{NS}(Q^2)$ we  obtain the quark--parton prediction
for the GLS sum rule. In order to obtain LO
and NLO expressions for the GLS sum rule one should
consider the NLO and NNLO approximations of the moments $M_n^{NS}(Q^2)$
correspondingly. Therefore, in order to make a self-consistent study
of the results of Table 1 for the GLS sum rule within the framework of
perturbative QCD, it is necessary to compare the results of the LO fit
with  the quark--parton expression of the GLS sum rule and the
results of the NLO fit with the LO expression of the GLS sum rule,
but with the coupling constant $\a_s$ defined through
Eq. (\ref{alnlo}), with $\L_{\overline{MS}}^{(4)}$ taken from the
results of the NLO fit.

Figures 1  demonstrate the experimental results for
the  GLS sum rule for both LO and NLO fits (see Table 1)
with the   statistical  experimental errors discussed above.
On the same Figures  we present the quark--parton and LO theoretical
expressions for the GLS sum rule (\ref{glsth}). In this work we are
neglecting the contributions of the HT corrections (which are known to
be quite important for the analysis of the GLS sum rules results in the
low-energy region \cite{BK,ChK,Ross}) since in the process of both our
fit  and of the one made by the CCFR group itself these corrections
were not taken into account in the expression for the SF $xF_3(x,Q^2)$.
\newpage
The values of the parameter
$\L_{\overline{MS}}^{(4)}$ in the  LO
perturbative QCD predictions depicted in Fig. 1b, namely
%\begin{equation}
%\L_{LO}=155 \pm 29 (stat) \pm 54    (syst)\ MeV;
%\label{LO}
%\end{equation}
\begin{equation}
\L_{\overline{MS}}^{(4)}=213 \pm 31 (stat) \pm 54 (syst)\ MeV
\label{LNLO}
\end{equation}
are taken in accordance with the results of our analysis of the
CCFR data for the SF $xF_3$ at the reference point $|Q_0^2|=3\ GeV^2$
(see Table 1 and discussions beyond it).
The statistical errors in Eq. (\ref{LNLO})  determine
the corresponding errors bars of the theoretical GLS sum rule predictions
(see Figs. 1).

It should be stressed that for the NLO fit the experimental values of
the GLS sum rule tend slowly to 3 from below, in qualitative agreement
with the theoretical expectations (see Fig. 1b). Moreover, within the
systematical experimental uncertainties our results are consistent with
the analysed perturbative QCD predictions.

However, at the
quantitative level  the tendency is toward the manifestation of
a certain disagreement between the perturbative QCD
predictions and the experimental results of  Table 1 obtained by us:
\begin{enumerate}
\item The results of Fig. 1a
demonstrate the slight $Q^2$-dependence  of the experimental
data for the GLS sum rule obtained in the process of the LO fit.
The obtained results lie below the quark--parton prediction
$GLS=3$.
\item
The NLO fit minimizes the disagreement presented at
Fig. 1a . However, even in this case the
discrepancy between the results of the NLO fit and
the LO QCD prediction for the GLS is surviving (see Fig. 1b)
\footnote{We have
checked that this disagreement does not disappear even after the
brute-force
inclusion of the NLO corrections into the theoretical expression
for the GLS sum rule.}.
The most surprising fact is that the minor discrepancy
takes place in the perturbative QCD region $|Q_0^2|>10\ GeV^2$

where we can safely follow our approximation of neglecting  the effects
of the HT contributions and TM corrections.
\end{enumerate}
\vspace{1mm}

{\bf 6. Discussion}

It seems to us that at the level of the statistical error bars
the results depicted in  Fig. 1b reveal
certain problems of  the explanation of
the experimental data for  the GLS sum rule within the framework of
the analysed QCD predictions. Indeed,  the LO theoretical QCD
expression of Eq. (\ref{glsth}) is approaching the asymptotic value
$GLS_{As}=3$ (which corresponds to the number of valence quarks inside
nucleon) somehow  faster than the results of our NLO fit of the
the CCFR experimental data.
In view of this  conclusion it is necessary to make several comments:
\begin{enumerate}
\item It seems problematic to describe the  deviation
that we observed between
theoretical and experimental results by taking into account threshold
effects, e.g. following the lines of the results of  recent studies
\cite{DSh}.
In the case of charged currents,
radiated by neutrinos,  the thresholds of production
of new flavours  manifest themselves in generations. Indeed,
the production of the $s$-quark is mainly accompanied by the production
of the $c$-quark in the whole region of momentum transfer.
Therefore we are taking $f=4$ in this region.
The mixing with
the quarks from the third generation are damped by the small values
of the Kobayashi--Maskawa matrix elements $V_{sb},V_{cb}$ and $V_{ct}$.
\item In accordance with the
discussions presented above, $b$- and $t$-quarks
appear simultaneously in the processes with
charged currents.
In the neutrino--nucleon deep-inelastic scattering, their contributions
should be studied in the region of very high $Q^2$.
\item The  deviation  of the experimental result
(\ref{result}),  the corresponding statistical uncertainties
taken into account, from the pure perturbative QCD
 predictions for the GLS sum rule,
with  $\Lambda_{\overline{MS}}^{(4)}$ defined by Eq. (\ref{La}),
was previously noticed even at the scale $|Q_0^2|=3\ GeV^2$
in the process of phenomenological
\cite{prep2,PHD} and theoretical \cite{ChK} studies. We  confirm this
observation and  stress that a similar inconsistency takes place
in a wide region of momentum transfer (see Figs. 1).

\item We can try to avoid this descrepancy by choosing
$\L_{\overline{MS}}^{(4)}$ as the free parameter and making a fit
of the experimental data presented in Fig. 1b
 on the GLS sum rule for
$|Q_0^2|<10\ GeV^2$. In this case we can describe the $Q^2$-behaviour
of the obtained experimental data satisfactorily using
$\L_{\overline{MS}}^{(4)}=724\pm153\ MeV$, which is too large to
support this procedure.
\item Notice once more that taking the HT contributions into account
\cite{BK,Ross} in the analysis can improve the agreement with the
theoretical predictions for the GLS sum rule at low $Q_0^2$
\cite{ChK,Ross}. However, at $|Q_0^2|>10\ GeV^2$ these corrections
cannot remove the observed  deviation between  the experimental
and theoretical results for the GLS sum rule.
\item It should be noted that in the region of small values of
$x$ ($x<0.015$)
not considered in the CCFR experiment, the more-complicated
 parametrization
of the SF can be used. The extrapolation of our simplest parametrization
(\ref{e10}) to this region can be a source of errors on the
experimental values of the GLS sum rule as calculated by us.
\item
It is worth mentioning  that  the results of the NLO fits
of the CCFR data for $\Lambda_{\overline{MS}}^{(4)}$, made with the help
of the  Altarelli--Parisi equation (see Eq. (\ref{La})) and using
Mellin moments (see Eq. (\ref{LNLO})), are in
exact agreement with the
central values from the results of  analogous  fits of the
$xF_3(x,Q^2)$ less precise data obtained
at Protvino \cite{Protvino}: $\Lambda_{\overline{MS}}^{(4)}=170\pm
60 (stat)\pm 120 (syst)\ MeV$ (AP) and $\Lambda_{\overline{MS}}^{(4)}=
230\pm 40 (stat)\pm 100 (syst)\ MeV$ (Moments).
\end{enumerate}
\vspace{2mm}
\newpage

{\bf 7. Conclusions}

In conclusion we would like to stress several points.
\begin{enumerate}
\item
Our NLO  result (\ref{LNLO})  for $\Lambda_{\overline{MS}}^{(4)}$ is in
agreement with the world average value of this parameter extracted from
the deep-inelastic scattering data \cite{Altarelli}.
\item
The obtained experimental $Q^2$-behaviour of the GLS sum rule is
consistent with the analysed perturbative QCD predictions,  within
systematical experimental uncertainties. However
at the level of the statistical experimental uncertainties
there is a certain discrepancy between theory and experiment.
\item
We do not exclude the possibility  that taking into account,
in our procedure, of the effects of the different QCD corrections,
namely of the $\a_s^2$ corrections to the coefficient function of
$xF_3(x,Q^2)$  \cite{vanNeerven},
the NNLO coefficients of the even anomalous dimensions
\cite{LRV}, still unknown NNLO
coefficients of odd anomalous dimensions and
of odd and even moments of $xF_3$, might remove the discrepancy
we found  between the experimental data for the GLS sum rule and
the corresponding theoretical prediction in the region of high-momentum
transfer. We hope that our work will push ahead the necessary
calculations and studies.
\item The third important problem is related to the necessity
of  experimental studies of the behaviour of the $xF_3(x,Q^2)$ SF
in the region of small $x$: $x<0.015$. Hopefully this problem can be
studied in the future at HERA, where it is planned to reach the region
$x\approx 10^{-4}$.
\end{enumerate}
\vspace{2mm}

%\newpage
{{\bf Acknowledgements}}

We are grateful to  M.H. Shaevitz and W.G. Selligman  for providing us
the CCFR data and to  J. Ch\'yla, G.T. Gabadadze and S.V. Mikhailov
for fruitful conversations.
We  wish to thank  W.  van Neerven and D.V. Shirkov for
discussions of the final results of our analysis.

We express our special thanks to G. Altarelli for useful comments
on the outcome of our studies.

The work of one of us (A.V.S.) is supported by the Russian Fund
of the Fundamental Research Grant N 94-02-04548-a.

\newpage
\begin{tabular}{|c|c|c|c|c|c|c|} \hline
\multicolumn{4}{|c|}{NLO} & \multicolumn{3}{c|}{LO}\\ \hline
$ |Q_0^2| $ & $\Lambda_{\overline{MS}}^{(4)}$ & $\chi^2_{d.f.}$ & GLS
& $\Lambda_{{LO}}^{(4)}$ & $\chi^2_{d.f.}$ & GLS  \\
$[GeV^2]$ & $[MeV]$ &
& sum rule   & $[MeV]$  &       & sum rule \\ \hline
  2 &  209 $\pm$32  &  71.5/62  &   2.401 & 154 $ \pm$16 & 87.6/62& 2.515 \\
  3 &  213 $\pm$31  &  71.5/62  &   2.446 & 154 $ \pm$29 & 87.7/62& 2.525 \\
  5 &  215 $\pm$32  &  71.8/62  &   2.496 & 154 $ \pm$28 & 88.0/62& 2.537  \\
  7 &  215 $\pm$34  &  72.2/62  &   2.525 & 155 $ \pm$27 & 88.3/62& 2.549  \\
 10 &  215 $\pm$35  &  72.6/62  &   2.553 & 154 $ \pm$29 & 88.5/62& 2.558  \\
 15 &  215 $\pm$34  &  73.2/62  &   2.583 & 155 $ \pm$28 & 88.8/62& 2.569  \\
 25 &  214 $\pm$31  &  74.1/62  &   2.618 & 155 $ \pm$17 & 89.2/62& 2.583  \\
 50 &  213 $\pm$33  &  75.4/62  &   2.661 & 155 $ \pm$27 & 90.2/62& 2.603  \\
 70 &  212 $\pm$34  &  76.1/62  &   2.680 & 155 $ \pm$26 & 90.3/62& 2.614  \\
 100&  211 $\pm$33  &  76.8/62  &   2.699 & 154 $ \pm$29 & 90.7/62& 2.623  \\
 150&  210 $\pm$34  &  77.6/62  &   2.720 & 154 $ \pm$29 & 91.2/62& 2.635  \\
 200&  209 $\pm$33  &  78.2/62  &   2.735 & 154 $ \pm$29 & 91.5/62& 2.643  \\
 300&  209 $\pm$33  &  79.0/62  &   2.755 & 153 $ \pm$29 & 92.0/62& 2.655  \\
 500&  207 $\pm$35  &  80.1/62  &   2.779 & 153 $ \pm$29 & 92.7/62& 2.664  \\
\hline
\multicolumn{7}{p{13cm}}{{\bf Table 1.} The results of the LO and NLO QCD fit
of the CCFR $xF_3$ SF data for  $f=4$, $|Q^2|>10\ GeV^2$, $N_{max}=12$
with the corresponding statistical errors. $\chi^2_{d.f.}$ is the
$\chi^2$ parameter normalized to the degree of freedom $d.f.$ }
\\[5mm]
\end{tabular}
\vspace{5mm}
\begin{center}
\begin{tabular}{|c|c|c|c|} \hline
  &\multicolumn{2}{c|}{ Our analysis}   & CCFR \cite{prep2,PHD} \\ \cline{2-4}
  &       LO        &     NLO           &  NLO                  \\ \hline
$A$ &6.86  $\pm$0.09  & 6.423 $\pm$0.088  & 5.976 $\pm$0.148\\
$b$ &0.795 $\pm$0.008 & 0.794 $\pm$0.012  & 0.766 $\pm$0.010\\
$c$ &3.38  $\pm$0.03  & 3.218 $\pm$0.035  & 3.101 $\pm$0.036\\ \hline
\multicolumn{4}{p{90mm}}{{\bf Table 2.} The parameters of quark distributions
$xF_{3}(x,Q_0^2)=Ax^{b}(1-x)^{c}$ at $|Q_0^2|=3\ GeV^2$.}\\[5mm]
\end{tabular}
\end{center}

\newpage

{\bf Figure captions}\\

\vspace{1cm}
Fig. 1a: The comparison of the results of the LO fit of the
$Q^2$ evaluation of the GLS sum rule with the statistical error bars
taken into account with the quark--parton prediction.

\vspace{1cm}
Fig. 1b: The comparison of the result  of the NLO fit of the
$Q^2$ evaluation of the GLS sum rule with the statistical error bars
taken into account with the LO perturbative QCD prediction.

\end{document}